# High-Sensitivity, High-Throughput Double Sagnac Lateral Shearing Quantitative Phase Microscopy and Tomography with Pseudo-Thermal Illumination

**Paweł Gocłowski[1], Maciej Trusiak[2], Balpreet S. Ahluwalia[1,3,4,†], Azeem Ahmad[1,*,†]**

[1]*Department of Physics and Technology, UiT The Arctic University of Norway, 9037 Tromsø, Norway*
[2]*Warsaw University of Technology, Institute of Micromechanics and Photonics, 8 Sw. A. Boboli St., 02-525 Warsaw, Poland*
[3]*Department of Clinical Science, Intervention and Technology, Karolinska Institute, Stockholm, Sweden*
[4]*The Faculty of Mathematics and Natural Sciences, Department of Physics, University of Oslo, 0313 Oslo, Norway*
[†]*Shared authors.*
**ahmadazeem870@gmail.com*
*Email: balpreet.singh.ahluwalia@uit.no, pawel.goclowski@gmail.com*

**Abstract:**

Quantitative phase microscopy (QPM) enables label-free measurement of local optical path length variations, providing critical insight into the structure and dynamics of transparent biological specimens. Here, a highly sensitive lateral shearing QPM (LS-QPM) system is presented, based on a novel double Sagnac common-path interferometric configuration combined with pseudo-thermal illumination. The pseudo-thermal light source plays a central role in enhancing spatial phase sensitivity by suppressing coherent noise and speckle artifacts, while maintaining sufficient temporal coherence to generate high-density interference fringes, thereby enabling robust single-shot phase retrieval. In addition, the double Sagnac architecture introduces an inherently stable common-path geometry, significantly enhancing temporal phase stability. Unlike conventional quadriwave lateral shearing interferometry (QWLSI)-based QPM systems, which typically suffer from a trade-off between spatial resolution and field of view (FOV), the proposed approach enables simultaneous achievement of diffraction-limited resolution and a large FOV. Experimental validation using calibrated polystyrene beads demonstrates accurate and spatially uniform phase reconstruction across the entire imaging area. Further, the system's capability for biological imaging is demonstrated through experiments on fixed and live HeLa cells, where subcellular features and dynamic processes are captured in a label-free manner. Furthermore, volumetric imaging of embedded bead samples highlights the potential of the approach for three-dimensional phase tomography. The lateral shearing mechanism, analogous to differential interference contrast (DIC), improves robustness to multiple scattering, indicating strong potential for future applications in imaging thick and heterogeneous samples.

## 1. Introduction:

Label-free optical microscopy has been recognized as an important tool for studying the dynamic behavior of living cells and their physiological processes [1-4]. In these techniques, the specimen is not stained or labeled, but instead it is visualized using intrinsic contrast mechanisms such as absorption, scattering, or refractive index. Among these, refractive index is particularly useful in cell biology because it offers the possibility of non-invasive imaging of transparent specimens in their native physiological state [4].

Techniques such as phase contrast microscopy (PCM) [5] and differential interference contrast (DIC) microscopy [6] take advantage of differences in optical path lengths, defined as the product of refractive index and physical thickness, to improve image contrast include. In PCM, phase differences are converted into detectable differences in intensity by introducing a phase difference between unscattered (background) and scattered light (carrying sample information) with the help of a phase plate placed at the Fourier plane of the objective lens. This technique helps to improve the contrast of weakly scattering structures and is extensively used for examining thin biological structures. However, PCM is plagued by halo artifacts and loss of contrast in thick structures, where multiple scattering prevents to extract accurate quantitative information.

DIC microscopy, on the other hand, offers excellent qualitative contrast by interfering two laterally sheared wavefronts. In this technique, a linearly polarized beam is divided by a Nomarski prism into two orthogonally polarized, spatially shifted beams that pass through the specimen with nearly identical scattering paths. This minimizes the scattering noise, allowing to image relatively thick samples, including tissues and embryos. The two beams, after passing through the second Nomarski prism and analyzer, create an interference pattern, offering high contrast differential images that highlight the local phase gradients. Even though this technique is robust for imaging thick samples, it is still qualitative in nature, without direct access to the quantitative phase information.

In order to overcome these limitations, quantitative phase microscopy (QPM) methods have been proposed for recovering the phase delay of light passing through transparent objects [4, 7-11]. QPM methods enable quantitative measurements of critical bio-physical parameters such as cell thickness, refractive index fluctuations, and dry mass density, thus enabling the study of sub-cellular dynamics using label-free methods [4, 7]. The performance of QPM methods is characterized by parameters such as spatial sensitivity, lateral resolution, temporal sensitivity, imaging speeds, field of view (FOV), and stability. Spatial sensitivity depends on the coherence properties of the illumination source [12]. In general, QPM methods using multiple frames for phase shifting achieve better spatial resolution at the cost of temporal resolution, whereas single-shot methods achieve fast acquisition at the cost of reduced spatial resolution. In addition, common-path interferometry methods are preferred for their high temporal stability, especially for applications involving long term imaging of dynamic biological specimens [13].

Extending the basic concepts of DIC and QPM, several techniques have been proposed to obtain quantitative phase information from DIC-type images. Of particular interest are Gradient Light Interference Microscopy (GLIM) [14, 15] and Quadri-Wave Lateral Shearing Interferometry (QWLSI) [16, 17]. These techniques allow the measurement of phase gradients, hence reconstructing the quantitative phase maps. GLIM, when applied on a DIC microscope, takes advantage of the inherent ability of DIC to suppress the effects of multiple scattering, hence applicable for imaging thick biological samples. However, it has the drawback of requiring the acquisition of four phase-shifted interferograms using a Spatial Light Modulator (SLM), reducing temporal resolution. The switching speed of liquid crystal-based SLMs also limits the acquisition rate, making GLIM less/not suitable for imaging fast dynamical processes. Further, the requirement of using a DIC microscope adds complexity, including the requirement for prism switching and alignment when changing the objective lens magnification. GLIM also lacks simultaneous intensity imaging, hence complicating the interpretation of the contrast mechanism, whether absorption or phase gradient related.

On the other hand, QWLSI utilizes a diffraction grating-based phase mask that creates multiple sheared replicas of the wavefront, allowing the acquisition of the gradient information of the phase in a single shot, followed by numerical reconstruction of the phase information QWLSI is a fast and simple technique, however it requires precise alignment of the phase mask with the camera pixels. Moreover, the phase mask is generally optimized according to the detector configuration, making the system costly and less flexible. Moreover, unwanted diffraction orders contribute to noise, compromising the accuracy of the technique.

The spatial resolution of QWLSI is limited by the pitch of the grating, causing a reduction of up to fourfold with respect to the diffraction-limited resolution of the microscope. Even though the resolution of the microscope can be increased to compensate for the reduction, it will compromise the field of view, effectively reducing the space-bandwidth product of the system.

These limitations include the loss of temporal resolution in GLIM and spatial resolution or FOV in QWLSI, thereby emphasizing the need to develop a quantitative phase imaging technique that is not only fast, robust, but also capable of diffraction-limited imaging with a large FOV.

In this work, a common-path lateral shearing quantitative phase microscopy (LS-QPM) system is presented. The proposed LS-QPM unit can be seamlessly retrofitted onto the output port of a conventional bright-field microscope, enabling quantitative phase imaging through lateral wavefront shearing. The common-path configuration ensures high temporal stability, while the single-shot implementation allows high-speed imaging. Importantly, the LS-QPM system achieves diffraction-limited spatial resolution over a large field of view, thereby overcoming the resolution–field-of-view trade-off inherent in QWLSI-based techniques. Furthermore, by employing DIC-like shearing principles without the need for dedicated DIC optics, the proposed approach provides robustness against multiple scattering, enabling reliable quantitative phase imaging of thick biological samples.

## 2. Material and Methods:

### 2.1. Mathematical description:

In the proposed four-beam interference configuration, four identical wave fronts are superimposed after shifting symmetrically around the optical axis. Consider $d_1$ and $d_2$ as the total shears along the $x$- and $y$-directions, respectively, such that each of the fields in a pair undergoes displacement $\pm d_1/2$ or $\pm d_2/2$, respectively. Two beams propagate with opposite tilt angles $-\theta_x$ and $+\theta_x$ along the $x$-direction, while the other two beams propagate with opposite tilt angles $-\theta_y$ and $+\theta_y$ along the $y$-direction. The corresponding complex fields can therefore be written as

$$E_1 = A_1 \exp\left[i\left\{\phi\left(x - \frac{d_1}{2}, y\right) - kx\sin\,\theta_x\right\}\right] \tag{1}$$

$$E_2 = A_2 \exp\left[i\left\{\phi\left(x + \frac{d_1}{2}, y\right) + kx\sin\,\theta_x\right\}\right] \tag{2}$$

$$E_3 = A_3 \exp\left[i\left\{\phi\left(x, y - \frac{d_2}{2}\right) - ky\sin\,\theta_y\right\}\right] \tag{3}$$

$$E_4 = A_4 \exp\left[i\left\{\phi\left(x, y + \frac{d_2}{2}\right) + ky\sin\,\theta_y\right\}\right] \tag{4}$$

where

$$k = \frac{2\pi}{\lambda}. \tag{5}$$

The recorded interferogram is given by

$$I = | E_1 + E_2 + E_3 + E_4 |^2. \tag{6}$$

Expanding Eq. (6) yields self-interference and cross-interference terms. Among these, the terms formed by the pairs $(E_1, E_2)$and $(E_3, E_4)$ are the ones of primary interest, since they encode the phase gradients along the $x$- and $y$-directions, respectively. Thus, the intensity may be written as

$$\begin{aligned} I \approx \quad & I_0 + 2A_1A_2\cos\left[\phi\left(x+\frac{d_1}{2},y\right)-\phi\left(x-\frac{d_1}{2},y\right)+2kx\sin\,\theta_x\right] \\ & +2A_3A_4\cos\left[\phi\left(x,y+\frac{d_2}{2}\right)-\phi\left(x,y-\frac{d_2}{2}\right)+2ky\sin\,\theta_y\right], \end{aligned} \tag{7}$$

where

$$I_0 = |A_1|^2 + |A_2|^2 + |A_3|^2 + |A_4|^2. \tag{8}$$

Assuming that the shear distances $d_1$and $d_2$are sufficiently small, the phase terms can be approximated using first-order Taylor expansion:

$$\phi\left(x+\frac{d_1}{2},y\right)-\phi\left(x-\frac{d_1}{2},y\right) \approx d_1\frac{\partial\phi}{\partial x}, \tag{9}$$

$$\phi\left(x,y+\frac{d_2}{2}\right)-\phi\left(x,y-\frac{d_2}{2}\right) \approx d_2\frac{\partial\phi}{\partial y}. \tag{10}$$

Substituting Eqs. (9) and (10) into Eq. (7) gives

$$I \approx I_0 + 2A_1A_2\cos\left(2kx\sin\,\theta_x + d_1\frac{\partial\phi}{\partial x}\right) + 2A_3A_4\cos\left(2ky\sin\,\theta_y + d_2\frac{\partial\phi}{\partial y}\right). \tag{11}$$

Using the carrier-frequency definitions

$$f_x = \frac{2\sin\,\theta_x}{\lambda}, f_y = \frac{2\sin\,\theta_y}{\lambda}, \tag{12}$$

Eq. (11) can be rewritten as

$$I \approx I_0 + 2A_1A_2\cos\left(2\pi f_x x + d_1\frac{\partial\phi}{\partial x}\right) + 2A_3A_4\cos\left(2\pi f_y y + d_2\frac{\partial\phi}{\partial y}\right). \tag{13}$$

For convenience, the modulation coefficients may be defined as

$$b_x = \frac{2A_1A_2}{I_0}, b_y = \frac{2A_3A_4}{I_0}, \tag{14}$$

so that the normalized intensity expression becomes

$$I \approx I_0\left[1 + b_x\cos\left(2\pi f_x x + d_1\frac{\partial\phi}{\partial x}\right) + b_y\cos\left(2\pi f_y y + d_2\frac{\partial\phi}{\partial y}\right)\right]. \tag{15}$$

Equation (15) shows that the recorded four-beam interferogram simultaneously contains two orthogonal carrier fringes, one modulated by the phase gradient along $x$ and the other by the phase gradient along $y$. Hence, the four-beam configuration enables simultaneous encoding of the differential phase information in both transverse directions within a single interferogram.

## 2.2. Experimental scheme:

Figure 3 shows the experimental realization of the proposed common-path lateral shearing quantitative phase microscopy (LS-QPM) system. The proposed system is based on an inverted bright-field microscope configuration, in which the pseudothermal (dynamic speckle [13, 18-20]) light source illuminates the specimen from the top. The scattered light collected from the specimen by the objective lens placed at the bottom of the stage is reflected by the folding mirror (M1) and then passes through the tube lens (L1), which forms an image of the specimen at the back focal plane, termed the image plane (IP).

The proposed system consists of an add-on unit that can be retrofitted at the output port of an existing bright-field microscope, thereby transforming it into an LS-QPM system. The QPM unit consists of various optical components, including polarizers, lenses, mirrors, and polarizing beam splitters, to form two orthogonal Sagnac interferometers. In the earlier versions of the QPM systems, a single Sagnac interferometer was used to obtain differential phase contrast information along only one direction [21]. Integration of the differential phase in the x-direction alone to obtain the phase information introduces line artifacts in the reconstructed images. However, in the proposed configuration, two orthogonal Sagnac interferometers are used to obtain phase gradients in both the x and y directions, thereby avoiding line artifacts in the phase images.

The image plane (IP) of the microscope is optically relayed to the back focal plane of lens L2 with a focal length of 200 mm. The output beam is first polarized at 45° using a polarizer (P1) so that both orthogonal polarization components are equally generated. Then, the beam is focused using lens L3 and directed towards a polarizing beam splitter (PBS1), which splits the beam into two orthogonally polarized beams. These two beams travel along different paths and are recombined at PBS1 using mirrors M3 and M4. This creates two sheared wavefronts with orthogonal polarization.

The recombined beams then pass through a second polarizer (P2), also oriented at 45°, which equalizes their polarization states and allows interference. Subsequently, the beam enters a second Sagnac interferometer, oriented orthogonally to the first. At PBS2, the beam is again split into orthogonal polarization components and recombined using mirrors (M5 and M6). As a result, this creates four sheared wavefronts that encode the phase gradients in both the x and y directions. These four beams are overlapped at the camera plane using lens L4 with a focal length of 200 mm. Before detection, a polarizer P3 ensures that all beams share the same polarization state, and creates an interferogram with a mesh pattern that encodes the orthogonal phase gradients. The shear magnitude and fringe spacing can be controlled by adjusting the angles of mirrors M3-M6, where M3-M4 control the shear along the x-direction and M5-M6 along the y-direction. The camera is mounted on a translation stage to fine-tune the shear between the overlapped wavefronts. It was found that systematic investigation of the shear amount is required for low noise and diffraction limited phase imaging.

The recorded interferogram is finally processed using a Fourier transform-based phase retrieval technique [22] to obtain the gradient phase maps in the two directions. Finally, the gradient maps are integrated along x and y axis to obtain the quantitative phase distribution of the specimen [23, 24].

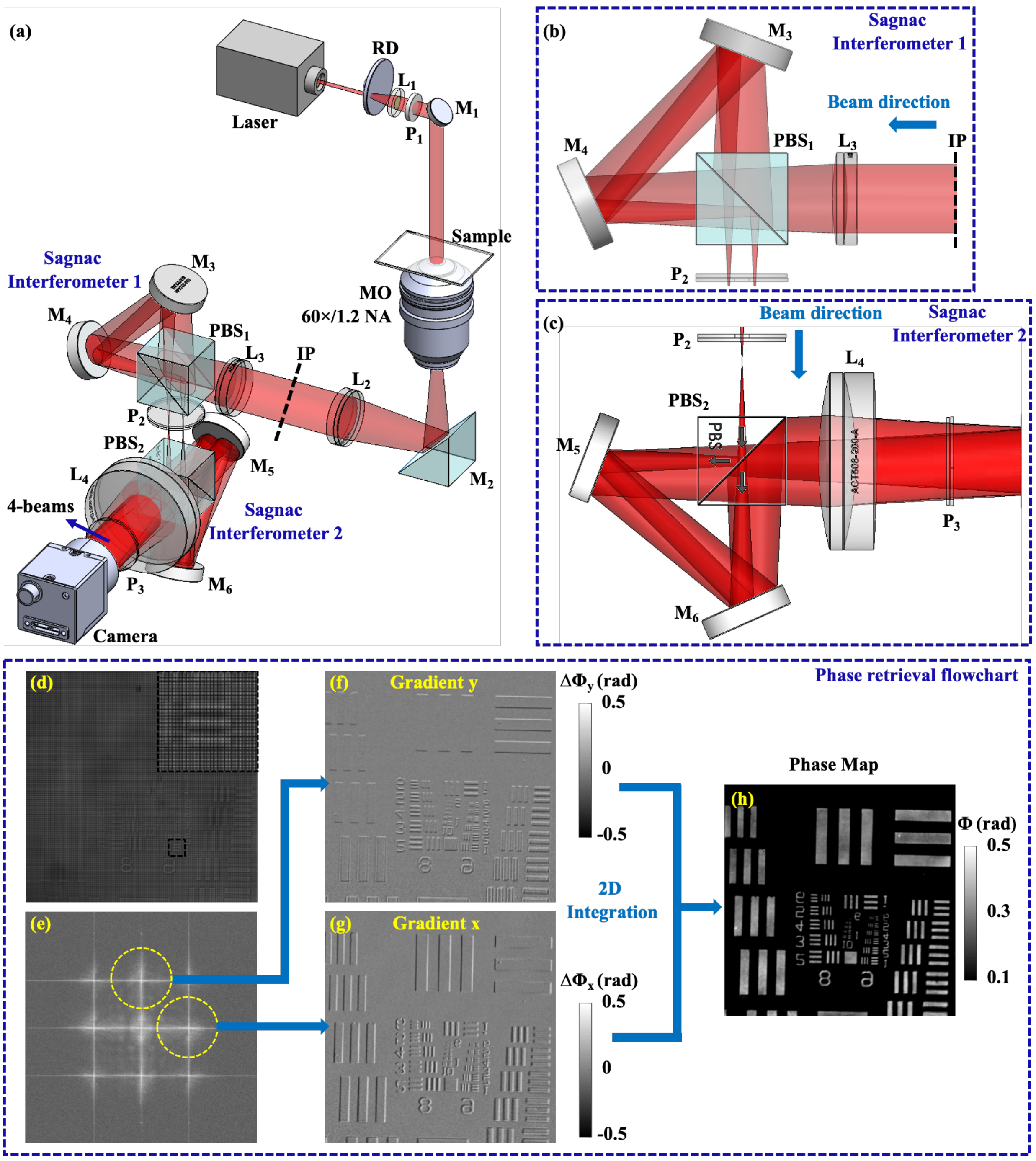

*Figure 1 – Schematic of the common-path lateral shearing quantitative phase microscopy (LS-QPM) unit. The system comprises two orthogonal Sagnac interferometers integrated at the output port of a bright-field microscope. The system uses polarizers (P1 – P3), polarizing beam splitters (PBS1, PBS2), and mirrors (M3, M4, M5, M6), as well as lenses (L2-L4), which generate and recombine four laterally sheared wavefronts, thus producing a mesh-like interferogram that encodes phase gradients along both x and y directions. Flowchart shows phase retrieval process: input grid interferogram with zoomed view, Fourier spectrum, differential phase in X axis, differential phase in Y axis and integrated phase map of 200 nm high phase USAF test.*

One of the main advantages of this configuration is that it has a common-path geometry, thus ensuring a high level of temporal phase stability and robustness to environmental vibrations. Additionally, the nearly zero optical path difference (OPD) between the interfering beams ensures high fringe contrast. Further, the partial shear operation of the LS-QPM suppresses the multiple scattering effects analogically to GLIM and enables quantitative phase imaging of thick and scattering biological specimens.

## 3. Results and discussion:

### 3.1. Spatial and Temporal Phase sensitivity:

Spatial and temporal phase sensitivities are key performance metrics for a QPM system. Spatial phase sensitivity quantifies the ability to detect small spatial variations in optical path length, whereas temporal phase stability reflects the stability of the measurement over time. To quantify these parameters, a sample-free interferogram sequence was recorded for 1 minute (Supplementary Video V1). The dataset consists of 1200 interferograms acquired at a rate of 20 Hz. This acquisition speed is sufficient to capture most relevant biological dynamics. Since the proposed technique operates in a single-shot manner, the imaging speed is primarily limited by the camera frame rate and can be increased to the kHz range (e.g., ~1 kHz) by using a faster camera.

Figure 2 presents the results of the phase sensitivity analysis. The grid interferogram and its corresponding Fourier spectrum are shown in Figs. 2(a, e). The Fourier peaks along the x- and y-directions were isolated using a circular mask and then inverse Fourier transformed to obtain gradient phase maps along both directions. Representative phase maps are shown in Fig. 2(b, f). The mean absolute value of these maps was calculated and used as a measure of the spatial phase sensitivity. Next, a phase map was computed for each frame and stacked to form a 3D matrix, where the xy plane represents the spatial dimensions and the z-axis represents time. Figure 2(c, g) shows the temporal phase fluctuation profile at a representative pixel of the recovered phase map. To quantify the temporal phase stability from the differential phase maps along x and y, the temporal phase profiles of all pixels were first averaged to obtain a global phase trend. This global profile was then subtracted from the temporal phase profile of each pixel. This step removes errors in the measured temporal stability that arise from global drift in the interferometric module. Finally, the standard deviation of the residual temporal phase signal was calculated for each pixel. These values were mapped to generate temporal phase stability maps, as shown in Fig. 2(d, h).

The spatial phase sensitivity was found to be 11.7 mrad along the x-direction and 13.3 mrad along the y-direction. The temporal phase sensitivity of the proposed LS-QPM system was measured to be 8.7 mrad along the x-direction and 9.7 mrad along the y-direction. This represents a substantial improvement compared to typical non-common-path systems, which generally exhibit phase stability on the order of ~30 mrad. The temporal stability map, however, reveals noticeable spatial variation across the field of view. For the x-direction, the variation appears predominantly along the horizontal axis, while for the y-direction, it follows the vertical axis. This trend seems to correlate with the orientation of the interferometric fringes. The underlying cause of this behavior is not yet fully understood and will be investigated in future work.

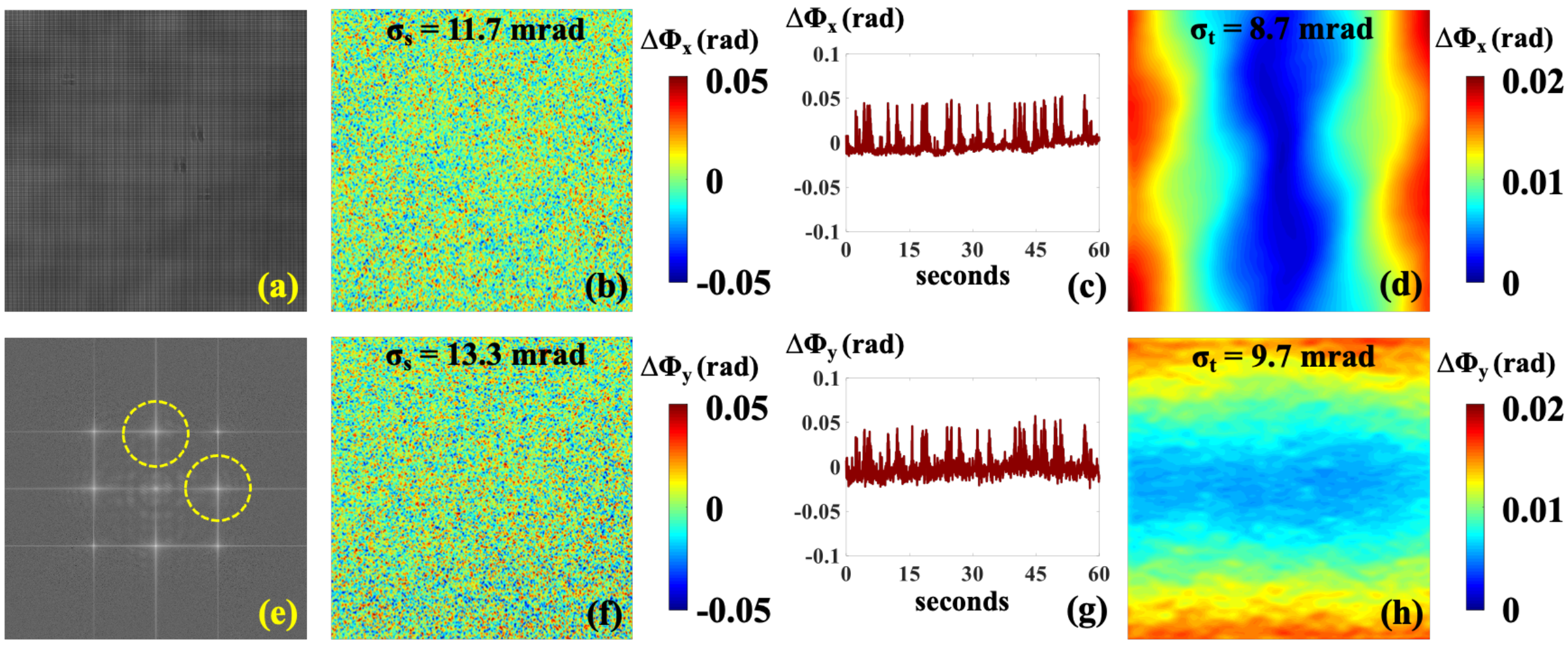

*Figure 2 – Spatial and temporal phase sensitivity investigation, based on 60-second-long sample-free movie. Input interferogram (a). Fourier spectrum (e). Differential phase in X and Y axis for single pixel as a function of time (c, g). Exemplary differential phase map in X and Y axis for spatial phase sensitivity investigation (b, f). Temporal phase sensitivity map for differential phase in X and Y axis (d, h).*

### 3.2. Shear investigation

The LS-QPM system works on the principle of partial lateral shearing interferometry. First, a systematic study was carried out to determine the optimal shear amount that enables accurate and low-noise phase reconstruction. It is observed that the spatial resolution and reconstruction quality depend strongly on the magnitude of shear between the interfering wavefronts. Experiments were performed using a sample of 1 μm beads, including regions with bead clusters. The analysis was repeated across different fields of view to ensure consistency. Figure 3 illustrates how the reconstruction quality of bead clusters varies with shear amount. Figure 3(a) represents the recovered phase map of the bead sample at a fixed shear value. The zoomed views of a region marked with yellow dotted box for different shear values are illustrated in Figs. 3(a1 – a9). To generate different shear values, the camera was mounted on a translation stage and translated between 3.0 mm and 7.0 mm in a step of 0.5 mm, as measured on the micrometer screw scale. The zero-shear position lies within this range. A broader analysis over a wider range is provided in the Supplementary Material (Fig. S1).

At camera positions of 3.0 – 4.0 mm (Fig. 3(a1–a3)), which correspond to shear values larger than the optimal range, the reconstructed images exhibit poor resolution, and individual beads cannot be clearly distinguished. This degradation arises because the amount of shear exceeds the effective diffraction-limited sampling of the system, leading to loss in the spatial resolution. At intermediate camera positions of 4.5–5.0 mm (Fig. 3(a4–a5)), the reconstruction quality improves significantly. The beads are well resolved, their boundaries are clearly defined, and the phase reconstruction fidelity is enhanced. However, further translation of the camera results in a gradual decline in image quality. At a position of 5.5 mm (Fig. 3(a6)), a noticeable loss of structural detail is observed, indicating information loss due to reduced shear. As the camera position approaches 6.0 mm (Fig. 3(a7)), the phase contrast is significantly diminished, and most structural features are suppressed. This indicates that the system is approaching the zero-shear condition, where the phase contributions from the overlapping wavefronts, including the sample information, nearly cancel out, preventing reliable phase recovery. Note that this position is not identical for differential phase measurements along the x- and y-directions. From the differential phase maps, the zero-shear position is estimated to be approximately 5.7 mm for the x-direction and 6.4 mm for the y-direction. The integrated phase map shown in Fig. S1 indicates a zero-shear position near 6.0 mm, which is approximately the average of the zero-shear positions obtained from the differential phase maps along the x- and y-directions. It is observed that the camera position at 5.0 mm is found to be optimal for the lossless and noise free phase recovery of the beads as shown in Fig. 3(a5). The selected optimal position at 5.0 mm corresponds to offsets of 0.7 mm and 1.4 mm from the zero-shear condition along the x- and y-directions, respectively. This discrepancy is not intrinsic to the method but is likely caused by residual optical aberrations, particularly distortion introduced by the tube lens.

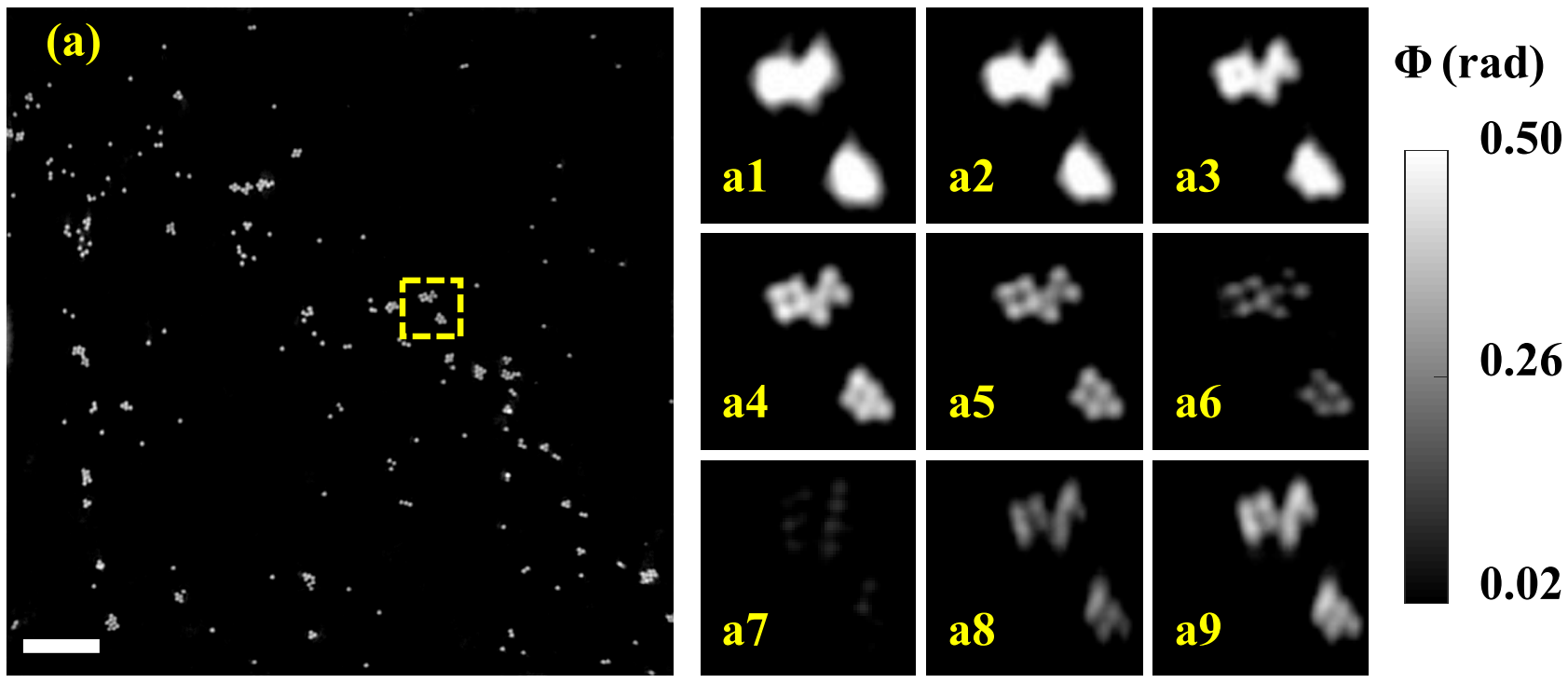

*Figure 3 – Shear investigation of bead clusters. Integrated phase map of 1 μm polystyrene beads is shown in (a). Panels (a1–a9) present zoomed views of two bead clusters at nine camera positions ranging from 3.0 to 7.0 mm in 0.5 mm increments. The scale bar corresponds to 20 μm.*

### 3.3. Phase reconstruction accuracy using standard bead samples:

The performance of the LS-QPM system was evaluated using standard polystyrene beads of two sizes: 1 μm and 300 nm. The accuracy of the reconstructed phase was assessed at multiple locations across the field of view. Figures 4(a, c) show the recorded grid interferograms for the two bead samples, while the corresponding reconstructed phase maps are presented in Figs. 4(b, d). All results correspond to a camera position of 5.0 mm, which was identified as the optimal shear condition. The reconstructed phase maps demonstrate that the LS-QPM system is capable of producing high-quality and well-resolved phase images for both bead sizes. It is important to note that phase recovery in LS-QPM involves a two-dimensional integration of differential phase maps. The accuracy of this integration step depends on the magnitude of shear along both the x- and y-directions. Ideally, equal shear values along both directions are required to ensure accurate phase reconstruction. However, in the present system, slight differences in shear along the x- and y-directions are observed for a fixed camera position due to the distortion present in the system. To understand the impact of this mismatch, the effect of varying shear ratios during the 2D integration process was systematically investigated. The results of this analysis are presented in Supplementary Fig. S2. Alternative approaches that rely on a single differential phase map typically require one-dimensional integration, which is prone to reconstruction artifacts. In Fig. 3 and 4, the color scale has been adjusted for better visualization. It is set to start from 0.02 rad to suppress background variations while retaining relevant sample features. Although the beads are successfully reconstructed across the entire field of view, variations in the maximum phase values and deviations from ideal spherical shape are observed. These differences are evident in panels (b1 – b4) and (d1 – d4) of Fig. 4. Such inconsistencies arise from spatial variations in the shear magnitude, as discussed in detail in the Supplementary section.

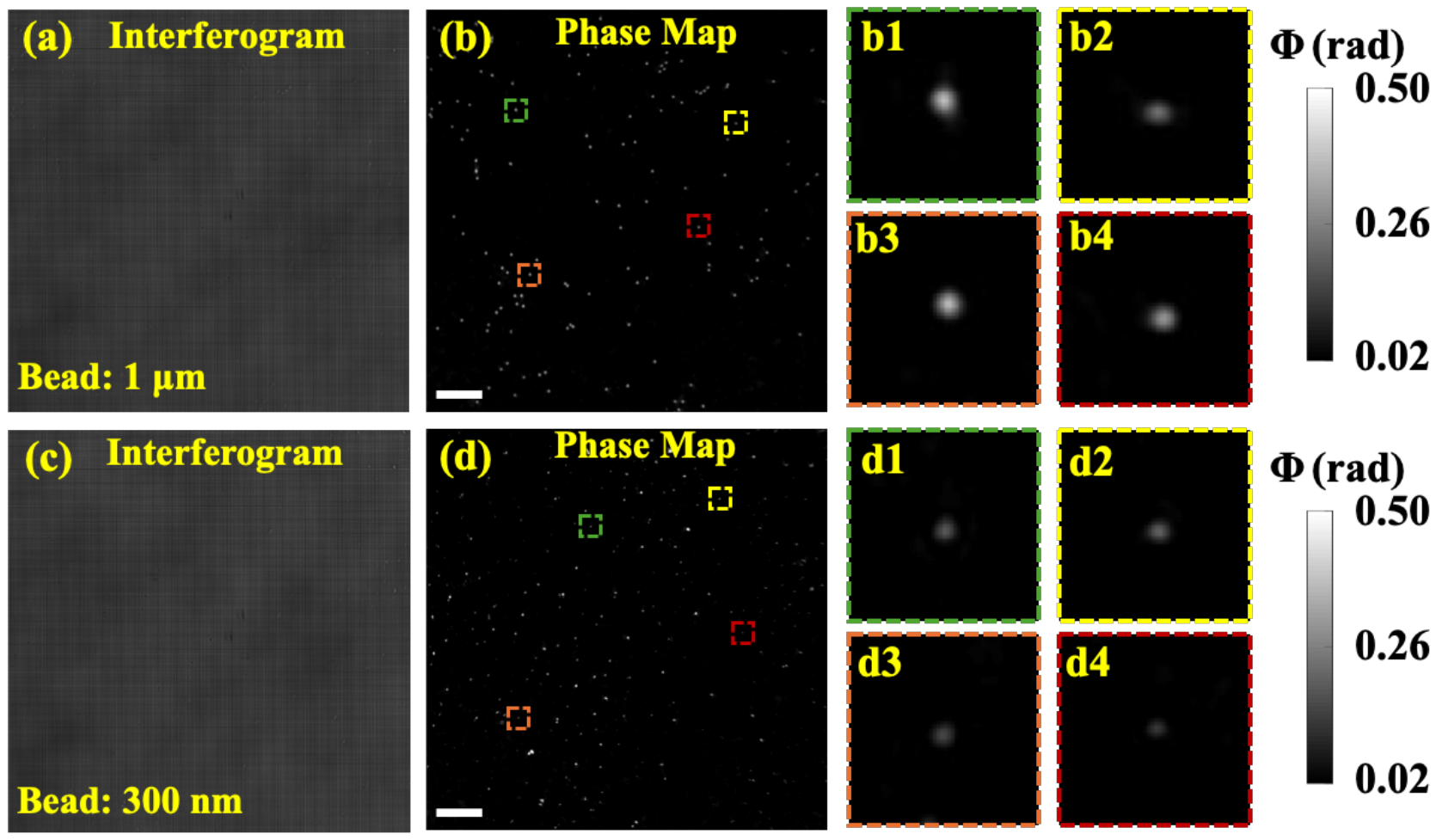

*Figure 4* – Phase imaging of polystyrene beads. Interferograms (a, c) and integrated phase maps (b, d) of 1 μm and 300 nm beads, respectively. Zoomed views of exemplary beads (b1-b4, d1-d4). Scale bar is equal to 20 μm.

### 3.4. Quantitative phase imaging of fixed HeLa cells:

The performance of the LS-QPM system was further evaluated using fixed HeLa cells. Figures 5(a, c) show the recorded grid interferogram and its corresponding Fourier spectrum. The Fourier peaks along the x- and y-directions were selected (indicated by yellow dotted circles) to reconstruct the gradient phase maps, as shown in Figs. 5(b, d). The final quantitative phase map obtained through 2D integration is presented in Fig. 5(e). The results demonstrate that the system provides high-quality phase reconstruction for biological samples. Both differential phase maps and the integrated phase map offer complementary information. The gradient phase maps enhance structural contrast by highlighting edges and provides quantitative differential phase information, while the integrated phase map enables true quantitative analysis of cellular properties.

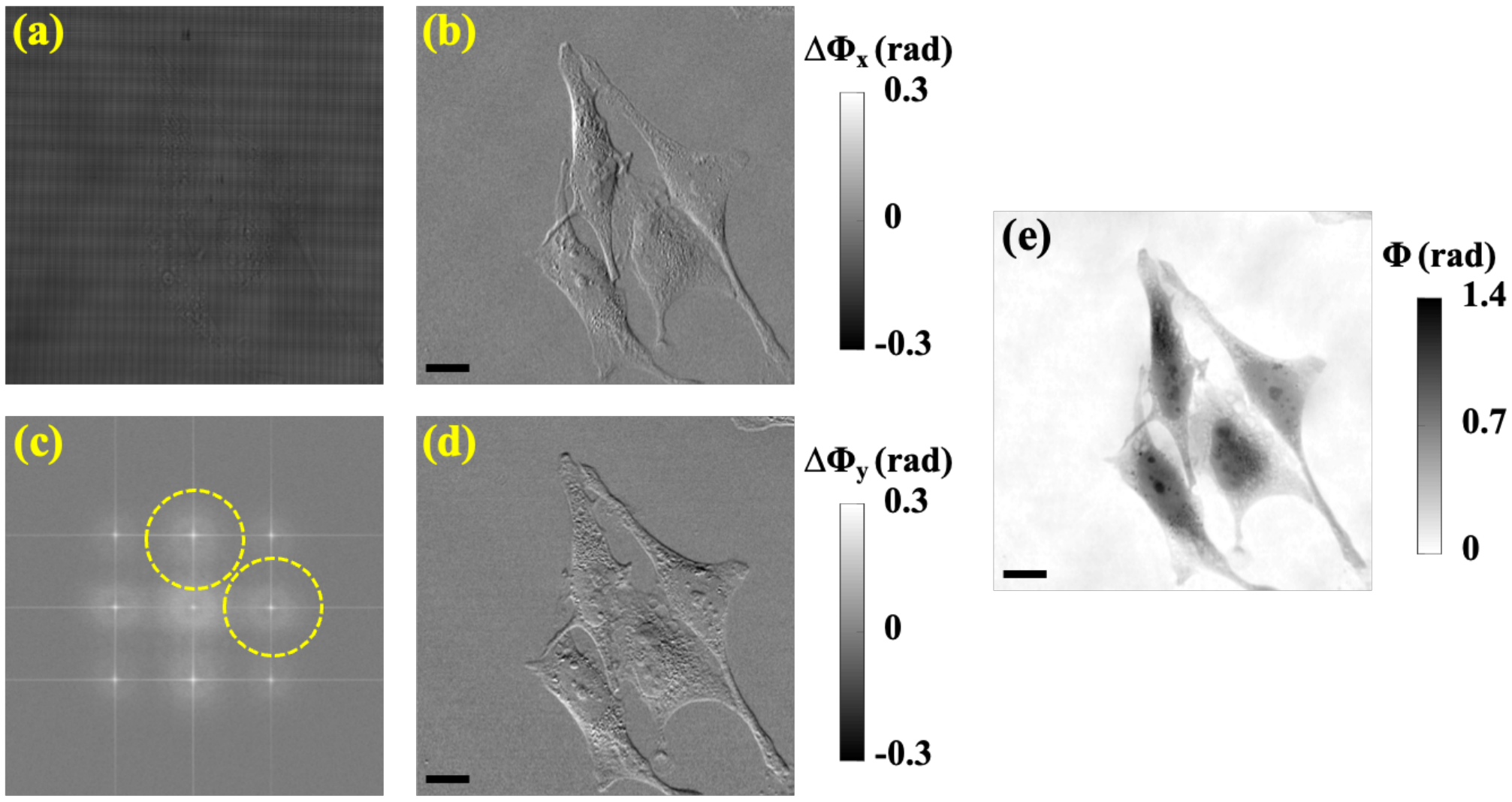

*Figure 5 – Phase imaging of fixed HeLa cells using LS-QPM system. Raw grid interferogram (a), Fourier spectrum (c), gradient phase map in X-axis (b), gradient phase map in Y-axis (d), integrated phase map (e). Scale bar is equal to 20 μm.*

An additional advantage of the proposed approach is the reduced need for phase unwrapping compared to conventional QPM, which require phase unwrapping if the phase of the specimens exceeds 2π limit. The measured differential phase values range between − 0.3 rad and + 0.3 rad, which is well below the ± π limit. This ensures that phase ambiguities are avoided under typical conditions, and only significantly thicker samples with abrupt phase variations may require unwrapping.

### 3.5. Quantitative phase imaging of live HeLa cells:

To further demonstrate the applicability of the LS-QPM system for dynamic biological studies, a time-lapse experiment was performed on live HeLa cells over a duration of 2 hours. To enable continuous monitoring of cellular dynamics, interferograms were acquired at 1 min intervals. The complete datasets are provided in the Supplementary Material, where Supplementary Video V2 shows the raw interferogram sequence and Supplementary Video V3 presents the corresponding reconstructed phase maps. Figure 6(a) illustrates representative integrated phase map of live HeLa cells. Panels a1-a9 in Figure 6 present a time-lapse sequence of a single cell at 15-minute intervals. These results highlight the capability of the system to capture temporal changes in cellular morphology and internal dynamics. During the experiment, the cells were imaged at room temperature outside standard incubation conditions, which may have induced cellular stress. Under these non-ideal conditions, the LS-QPM system successfully captured progressive morphological changes, including gradual contraction of the cell membrane and active intracellular motion. These observations demonstrate the strong potential of the technique for long-term, label-free monitoring of live-cell behavior with quantitative phase information. When combined with a controlled live-cell incubation environment, the system can enable stable and physiologically relevant imaging over extended durations, facilitating detailed studies of cellular dynamics and function.

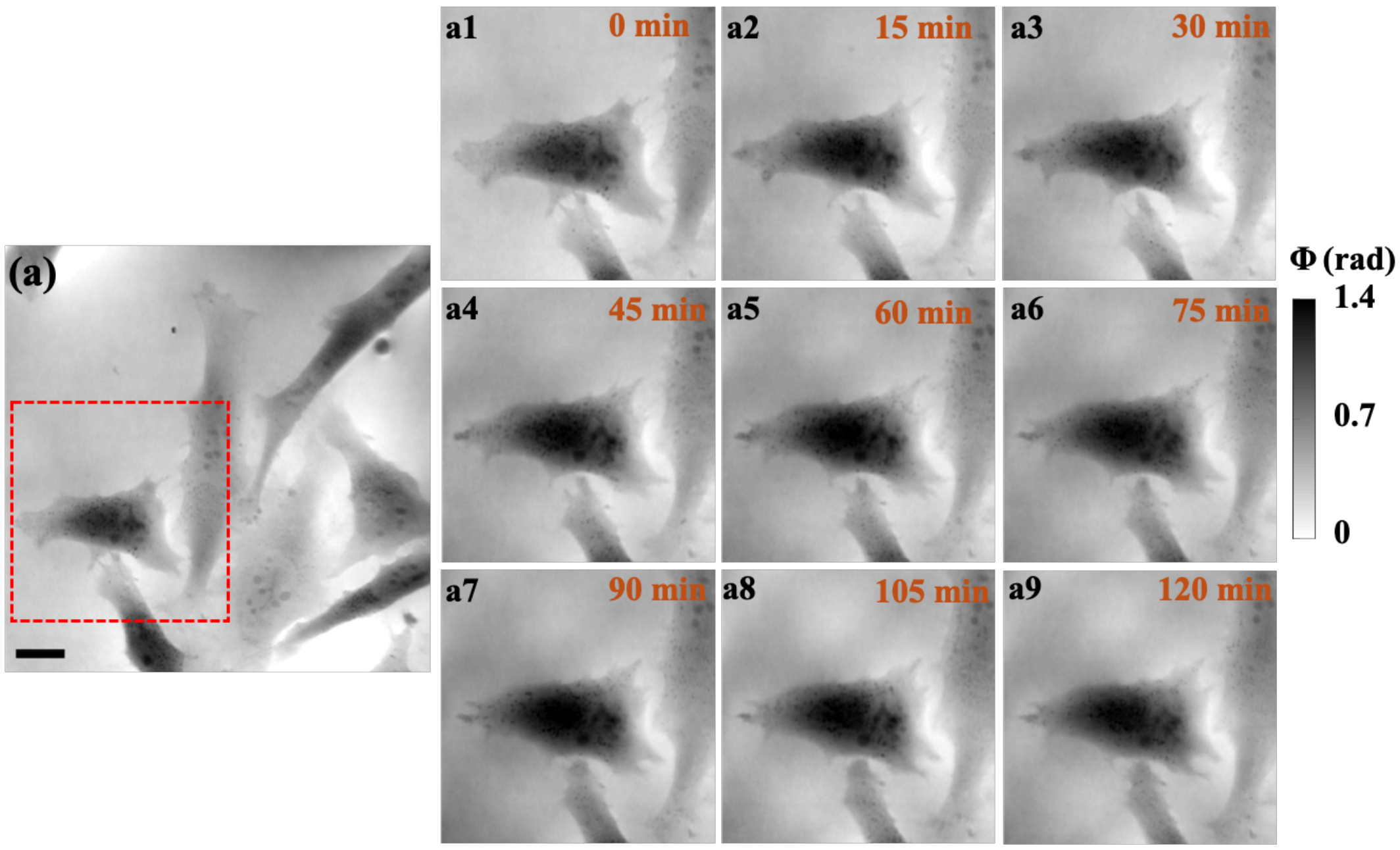

*Figure 6 – Timelapse of live HeLa cells. Integrated phase map of the entire FoV (a). Panels (a1-a9) show zoomed view of singular HeLa cell at 9 consecutive timestamps: 0, 15, 30, 45, 60, 75, 90, 105 and 120 min. Scale bar is equal to 20 µm.*

### 3.6. Quantitative phase tomography of volumetric bead sample:

Finally, the capability of the LS-QPM system for three-dimensional phase tomography was investigated. A volumetric sample composed of 1 µm polystyrene beads embedded in an approximately 100 µm thick polydimethylsiloxane (PDMS) layer was used for this study. Interferograms were recorded while scanning the sample along the axial direction with a step size of 0.5 µm. A representative single-plane phase map is shown in Fig. 7(a). Orthogonal cross-sectional views along the X–Z, and Y–Z planes, taken along the indicated yellow dashed lines, are presented in Figs. 7(b, c). The reconstructed three-dimensional integrated phase volume is shown in Fig. 7(d). The isometric rendering demonstrates that the beads are consistently reconstructed and localized throughout the sample depth. Further, the axial resolution of the system was estimated using the full width at half maximum (FWHM) of the bead line profile along the z-direction. This analysis was performed on a line profile extracted from the X–Z cross-section (Fig. 7(b)). The measured FWHM is approximately 9 pixels, corresponding to an axial resolution of ~ 4.5 µm. The ability of the LS-QPM system to perform phase-resolved volumetric imaging with high phase sensitivity underscores its potential for three-dimensional quantitative imaging applications. Note that the axial resolution of the proposed system can be further improved by incorporating oblique illumination in future implementations. In addition, the approach can be extended to optical diffraction tomography if multiple interferograms corresponding to different illumination directions are acquired and used for 3D refractive index reconstruction.

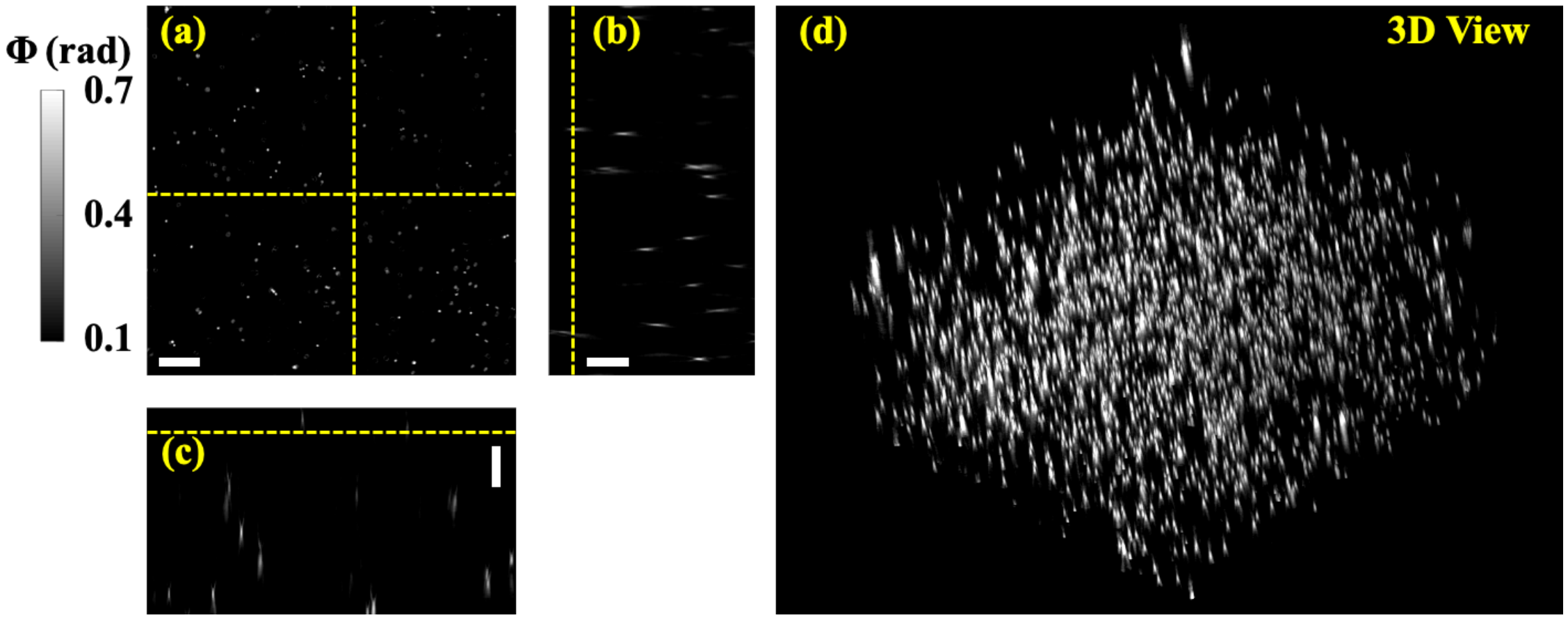

*Figure 7 – Z-scanning through 100 µm PDMS layer with embedded 1 µm polystyrene beads. Cross-sections X-Y (a), X-Z (b) and Y-Z (c). Isometric view of the entire volume (d). Scale bars are equal to 20 µm.*

## 4. Conclusion:

In this work, a common-path lateral shearing quantitative phase microscopy (LS-QPM) system has been developed and experimentally validated as a stable and versatile platform for quantitative phase imaging. The proposed module can be readily integrated into a conventional bright-field microscope, allowing straightforward implementation without significant modification of the existing optical setup. Owing to its common-path design, the system exhibits high temporal stability and reduced sensitivity to environmental perturbations. In addition, the single-shot acquisition scheme enables high-speed imaging, with performance primarily limited by the camera frame rate. The system provides diffraction-limited spatial resolution over a relatively large field of view, mitigating the conventional trade-off in common-path QPM systems between resolution and imaging area. The DIC-like lateral shearing mechanism further enhances structural contrast and improves robustness in the presence of multiple scattering, making the approach well suited for thick and heterogeneous samples. A systematic investigation of shear dependence indicates the presence of an optimal shear range, which is critical for achieving accurate and low-noise phase reconstruction.

Both spatial and temporal phase sensitivities were experimentally characterized, confirming improved stability compared to non-common-path interferometric configurations. The performance of the system was further validated using standard polystyrene bead samples, demonstrating consistent and reliable phase recovery across the field of view. The applicability of LS-QPM to biological imaging was demonstrated using both fixed and live HeLa cells. The system successfully resolved subcellular structures and captured dynamic behavior in a label-free manner, highlighting its suitability for long-term live-cell studies. Furthermore, volumetric phase imaging of embedded bead samples demonstrates the potential of the approach for three-dimensional phase tomography. While the current axial resolution remains moderate, it can be further enhanced through the use of oblique illumination and multi-angle acquisition strategies. The proposed LS-QPM system combines temporal stability, high spatial phase sensitivity, and high-speed quantitative imaging, and shows strong potential for future developments in three-dimensional quantitative microscopy.

**Data availability statements.** The datasets generated and/or analyzed during the current study are available from the corresponding author upon reasonable request.

**Funding.** A.A. acknowledges FRIPRO Young (project # 345136) funding from Research Council of Norway. B.S.A. and P.G. acknowledges the funding from UiT Thematic Funding (NASAR). B.S.A. acknowledges support from the UiT Talent Innovation Grant, Research Council of Norway (Verification Project: 360778) and European Union's HORIZON Research and Innovation Actions under grant agreement No. 101191315. M.T. acknowledges project No. WPC3/2022/47/INTENCITY/2024 funded by the National Center for Research and Development as part of the 3rd competition for joint research projects as part of Polish-Chinese cooperation (2022). The publication charges for this article have been funded by a grant from the publication fund of UiT The Arctic University of Norway.

**Author Contribution.** A.A. conceptualized the idea and designed the optical configuration. A.A. and B.S.A supervised and conceived the project. P.G and A.A. developed the experimental setup. P.G. performed the experiments, analyzed the data, and prepared the figures. A.A. and P.G. mainly wrote the manuscript. M.T. contributed to the insights into LS-QPM. All authors reviewed and edited the manuscript.

**Conflict of interest.** Azeem Ahmad and Balpreet Singh Ahluwalia have submitted a patent application to protect the invention of LS-QPM (Patent Application No. 2513325.7). All other authors declare no competing interests

**Acknowledgements.** We would like to acknowledge Dr. Hong Mao for preparing fixed and live HeLa samples.

**Supplemental document.** See Supplement 1 for supporting content.

# Supplementary Material

## High-Sensitivity, High-Throughput Double Sagnac Lateral Shearing Quantitative Phase Microscopy and Tomography with Pseudo-Thermal Illumination

**Paweł Gocłowski[1], Maciek Trusiak[2], Balpreet S. Ahluwalia[1,3,4,†], Azeem Ahmad[1,*,†]**

[1]*Department of Physics and Technology, UiT The Arctic University of Norway, 9037 Tromsø, Norway*
[2]*Warsaw University of Technology, Institute of Micromechanics and Photonics, 8 Sw. A. Boboli St., 02-525 Warsaw, Poland*
[3]*Department of Clinical Science, Intervention and Technology, Karolinska Institute, Stockholm, Sweden*
[4]*The Faculty of Mathematics and Natural Sciences, Department of Physics, University of Oslo, 0313 Oslo, Norway*
[†]*Shared authors.*
**ahmadazeem870@gmail.com*
*Email: balpreet.singh.ahluwalia@uit.no, pawel.goclowski@gmail.com*

**Visualization 1:** Time-lapse sample free interferometric recording for spatial and temporal phase stability.

**Visualization 2:** Time-lapse interferometric recording of live HeLa cells.

**Visualization 3:** Time-lapse quantitative phase map of live HeLa cells.

### 1. Shear investigation:

The selection of an appropriate shear value is critical in shear-based interferometric microscopy. In general, large shear values tend to degrade spatial resolution in the reconstructed phase, whereas very small shear reduces the signal-to-noise ratio. Therefore, an optimal intermediate shear must be identified to balance these competing effects.

In the proposed LS-ODT system, the shear between the overlapping wavefronts is controlled by adjusting the camera position using a one-dimensional translation stage with a total travel range of 12 mm. The stage was initially positioned at the midpoint (6 mm), corresponding approximately to the zero-shear condition, and securely fixed to the optical table. Subsequently, the full translation range from 0 to 12 mm was systematically scanned to determine the optimal shear. A total of 25 interferograms were acquired using 1 µm and 300 nm polystyrene beads, covering the range from 0 mm to 12 mm in steps of 0.5 mm. At each position, the sample was refocused using a piezo stage to ensure accurate shear characterization, as camera translation introduces defocus in the image. Figure S1 presents the full field-of-view integrated phase maps, along with zoomed views of a single bead for all camera positions.

The results reveal consistent trends across both bead sizes and across the entire translation range. At the lower end of the range (e.g., 0.0 – 2.5 mm), the shear is excessively large, leading to significant distortion of the bead shape. In the intermediate range (approximately 3.0 – 5.0 mm), the bead shape is generally preserved; however, noticeable variations indicate a loss of spatial resolution. As the shear approaches zero (for camera positions 5.0 – 6.0 mm), the phase contrast progressively diminishes. While the bead remains well defined at 5.0 mm, it becomes increasingly distorted at 5.5 mm and is no longer discernible at 6.0 mm. Based on these observations, a camera position of 5.0 mm was initially selected as an optimal operating point, providing a reasonable balance between spatial resolution and phase sensitivity. This estimate was obtained from the coarse scan presented in Fig. S1, which used a step size of 0.5 mm to cover the full translation range. To refine the optimal position, a detailed study was performed using a phase USAF resolution target over the 4.5–5.5 mm range with a step size of 0.1 mm, as described in Section 3.

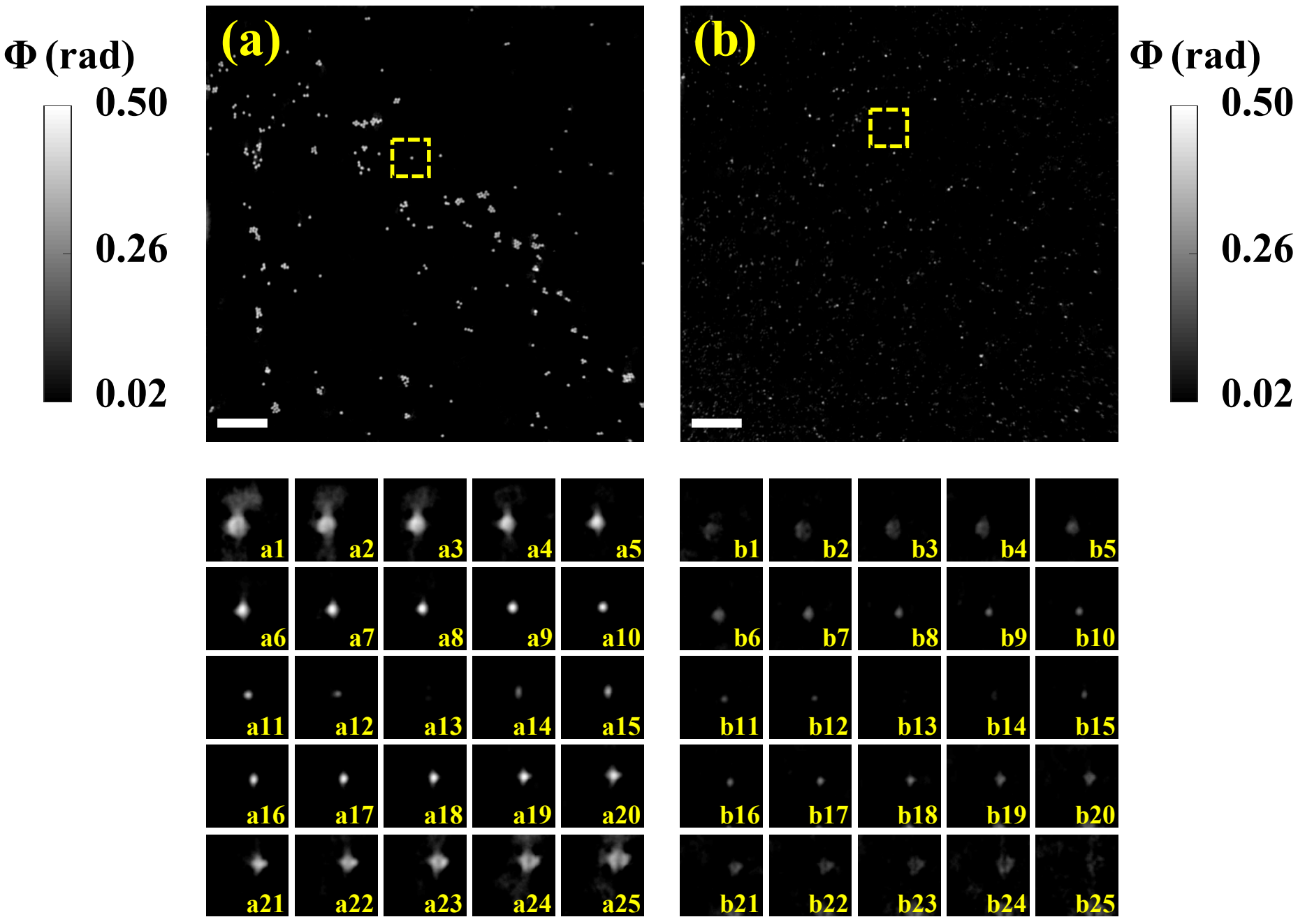

*Figure S1 – Shear investigation. Integrated phase maps of 1µm (a) and 300 nm polystyrene beads (b). Panels (a1-a25 and b1-b25) show a zoomed view of singular bead for 25 camera positions ranging from 0 to 12 mm in 0.5 mm increments. Scale bar is equal to 20 µm.*

## 2. Optimization of integration parameters (dx, dy) under anisotropic shear

In practical implementations, the shear introduced along the x- and y-directions is not always identical. As a result, using equal values for the integration parameters $dx$ and $dy$ during phase reconstruction may lead to inaccuracies. This effect was observed in the initial reconstructions, where identical values of $dx$ and $dy$ resulted in a systematic elongation of the reconstructed beads along one direction.

To address this issue, a systematic study was conducted to evaluate the impact of different $dx$: $dy$ ratios on the reconstructed phase. The results of this analysis are presented in Fig. S2, where the corresponding ratios are indicated above each panel (a1 – a9). It was found that a ratio of 1.5:1 (panel a3) provides the most accurate reconstruction, as evidenced by the improved sphericity of the bead. Deviations from this ratio resulted in noticeable anisotropic distortion in the reconstructed phase profiles. Based on these findings, all subsequent results presented in Figs. 4 – 7 were processed using the optimized $dx$: $dy$ ratio to ensure accurate and isotropic phase reconstruction.

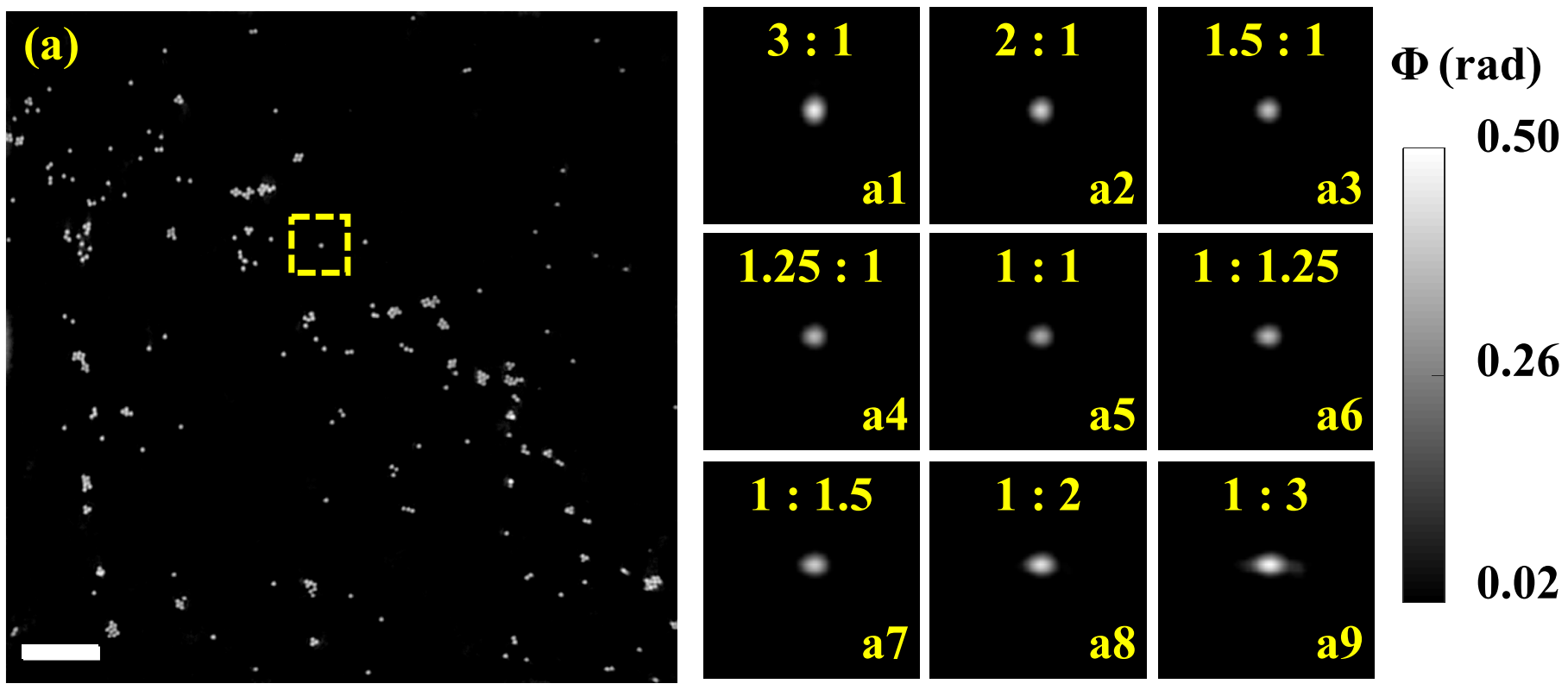

*Figure S2 – Phase integration analysis for varying $dx$:$dy$ ratios. (a) Integrated phase map of 1 µm polystyrene beads. Panels (a1–a9) present magnified views of a representative bead reconstructed using nine different $dx$:$dy$ ratios, indicated above each panel. Scale bar: 20 µm.*

## 3. Fine optimization of camera position using phase resolution target

The initial studies presented in Fig. S1 identified 5.0 mm as an approximate optimal camera position for the LS-QPM system. However, these measurements were performed with relatively coarse steps of 0.5 mm to cover the full translation range. To refine this estimate, a higher-resolution study was conducted using a phase USAF resolution target. Figure S3 shows the lateral resolution analysis over a narrower range (4.5 – 5.5 mm) with a step size of 0.1 mm. The results reveal a noticeable difference in resolving power along the x- and y-directions, which can be attributed to unequal shear in the two axes. For positions between 4.5 mm and 4.8 mm (panels a1 – a4), the fourth element of Group 10 is resolved along the x-axis, while the third element is resolved along the y-axis. In the range of 4.9 – 5.1 mm (panels a5 – a7), the fourth element is resolved in both directions. At 5.2 mm and 5.3 mm (panels a8 – a9), the fifth element becomes resolvable along the x-axis, indicating improved resolution. However, at positions beyond 5.3 mm, a significant loss of structural information is observed, particularly in the higher-frequency elements.

Although the raw resolution analysis might suggest 5.2 mm as the optimal position, resolution alone does not fully determine reconstruction quality. Across the 4.5 – 5.5 mm range, a gradual reduction in signal-to-background ratio and phase contrast is observed as the shear approaches zero. At 5.2 mm, the phase features of the USAF target begin to merge with the background, indicating reduced contrast. Furthermore, due to system distortions, the shear is not uniform across the field of view, leading to spatially varying degradation, most notably in the right region of the image, where even lower-resolution elements are no longer reliably resolved. To ensure uniform phase quality across the full field of view, a camera position of 4.8 mm was selected as the optimal operating point and used throughout this study. This choice represents a compromise, accepting a minor reduction in nominal resolution in exchange for improved contrast and uniformity. At the selected optimal position, the lateral resolution of the integrated phase map is approximately 690 nm along the x-axis and 776 nm along the y-axis. This is slightly worse than the theoretical resolution limit of the system, which is approximately 500 nm and may be improved in future implementations through the use of an ultra-low-distortion tube lens.

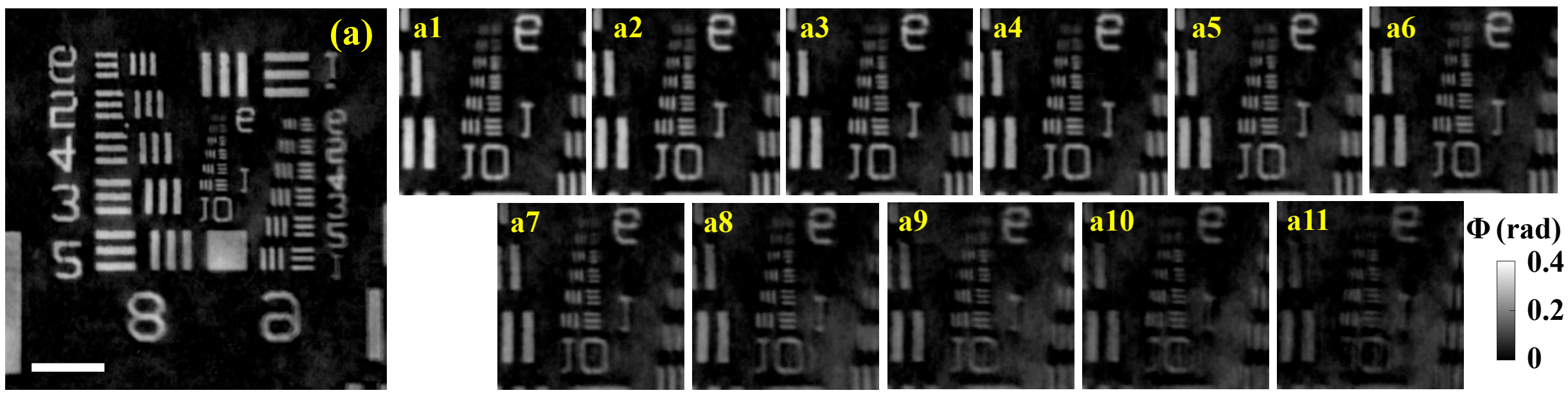

*Figure S3 – Fine resolution and shear analysis. (a) Integrated phase map of a USAF phase resolution target. Panels (a1–a11) present zoomed views of Group 10 for eleven camera positions (4.5 – 5.5 mm, 0.1 mm step). Scale bar: 20 µm.*